\documentclass[twocolumn,secnumarabic,amssymb, nobibnotes, aps, prd]{revtex4-1}

\usepackage{graphicx}
\usepackage{dcolumn}
\usepackage{bm}
\usepackage{slashed}

\begin{document}

\preprint{AIP/123-QED}

\title[XIV Hadron Physics, Florian{\'{o}}polis, Brazil, March 18-23 of 2018.]{Some diagrams and basic formalism of the \\low energy kaon-hyperon interaction}
\thanks{ Poster presented at the XIV Hadron Physics, 18th-23nd March
2018, in Florian{\'{o}}polis (Brazil)}

\author{M. G. L. N. Santos}
 \email{magwwo@gmail.com}
\author{C. C. Barros, Jr}%
 \email{barros.celso@ufsc.br}
\affiliation{ 
Departamento de F{\'{i}}sica, CFM, Universidade Federal de Santa Catarina\\ Florian{\'{o}}polis SC, CEP 88010-900, Brazil
}%

\date{\today}

\begin{abstract}
In this work the low energy kaon-hyperon interaction is studied with nonlinear chiral invariant Lagragians considering
kaons, hyperons, and the corresponding resonances in the intermediate states. We show the basic formalism to calculate the total cross sections,
angular distributions, and some diagrams of interest.
\end{abstract}

\maketitle

 \section{Introduction}
In this work we show the mathematical procedure to calculate the diagrams for the
low energy
kaon-hyperon ($KY$) interactions considering a model \cite{BarrosJr2001} that has been used in the study of the low energy
pion-hyperon ($\pi Y$) interactions. 
When studying low energy kaon interactions, obviously the processes have different features and 
different particles must be considered in the intermediate states.

 In this paper we just study the direct diagrams for each baryon 
resonance, and a complete calculation, taking into account the crossed diagrams
and other possible interactions will be published soon. 

We show how to calculate the total cross sections and
angular distributions for $K\Lambda$ and $K\Sigma$ interactions using nonlinear chiral invariant Lagragians in the center-of-mass frame.

 We also expect that these calculations allow us to calculate other potentials like the $NY$, $YY$ and $NNY$, which are very important in the studies of neutron stars and hypernuclei.

\section{The Method}

To study the $KY$ interactions we make an analogy with the $\pi N$ interaction that is very well studied, many models and a large amount of experimental data are available.
Here we shall use an effective chiral model for the
 $\pi N$ scattering and then extend it to the kaon-hyperon case.

The chiral Lagrangians \cite{Coelho1983} of the  $\pi N$ interaction with spin 1/2 and 3/2 baryons are given by 
\begin{equation}
\label{eq1}
\mathcal{L}_{\pi NN}= \frac{g}{2m}\big(\overline{N}\gamma_\mu\gamma_5\vec{\tau}N\big)\cdot\partial^\mu\vec{\phi},
\end{equation}
\begin{eqnarray}
\label{eq2}
\mathcal{L}_{\pi N\Delta}=  g_\Delta\bigg\{\overline{\Delta}^\mu\Big[g_{\mu\nu}-(Z+1/2)\gamma_\mu\gamma_\nu\Big]\vec{M}N\bigg\}.\partial^\nu\vec{\phi},
\end{eqnarray}
where $N$, $\Delta$, $\vec{\phi}$ are the nucleon, delta, and pion fields with masses $m$, $m_\Delta$ and $m_\pi$, respectively. $\vec{M}$ and $\vec{\tau}$ are isospin matrices, and $Z$ is a parameter representing the possibility of the off-shell-$\Delta$ having spin 1/2. The parameters $g$ and $g_\Delta$  are coupling constants and depend on each intermediary particle. 

We calculate the Feynman amplitudes and sum  all the  isospin contributions to obtain the  $T_{KY}$ amplitude, that is given by
\begin{equation}
 T_{KY}=\sum_IT^IP_I,
\label{eq5}
\end{equation}
with
\begin{equation}
 T^I=\overline{u}(\vec{p'})\Big[A^{I}+\frac{1}{2}(\slashed{k}+\slashed{k}')B^I\Big]u(\vec{p}),
\label{4}
\end{equation}
where the subscript $ (KY)$ represents the initial particles, a spin 0 kaon and a spin 1/2 hyperon. $P_I$ are the projector operators of total isospin states,
 $T^I$, the respective amplitudes, $u(\vec{p})$ is a spinor 
representing the initial baryon with $\vec{p}$ momentum and $k$ is the meson four-momentum.
So we can calculate the $A_I$ and $B_I$ amplitudes for each total isospin channel in the scattering.  
 
The scattering matrix for a given isospin state is
\begin{equation}
M^I=\frac{T^I}{8\pi \sqrt{s}}=G^I + H^I i \vec{\sigma}.\hat{n},
\end{equation}
which may be decomposed into the spin-non-flip and spin-flip amplitudes $G(\theta)$ and $H(\theta)$, and then expanded in partial-wave amplitudes
\begin{equation}
 G^I=\sum_{l=0}^\infty{\Big[(l+1)F_{l+}^I+lF_{l-}^I\Big] P_l (\theta)},
\end{equation}
\begin{equation}
 H^I=\sum_{l=1}^\infty{\Big[F_{l-}^I-F_{l+}^I\Big] P_l^{(1)} (\theta)}.
\end{equation}
The partial-wave amplitudes are, using the Legendre polynomial's orthogonality relations 
\begin{equation}
 F_{l\pm}^I=\frac{1}{2}\int_{-1}^1\Big[P_l(\theta)f^I_1(\theta) +P_{l\pm 1}(\theta)f^I_2(\theta)\Big] d\theta,
\end{equation}
where	
\begin{eqnarray}
 f_1^{I}(\theta)=\frac{(E+m)}{8\pi \sqrt{s}}[A^{I}+(\sqrt{s}-m)B^I],
\end{eqnarray}
\begin{equation}
 f_2^{I}(\theta)=\frac{(E-m)}{8\pi \sqrt{s}}[-A^{I}+(\sqrt{s}+m)B^I],
\end{equation}
where $E$ is the hyperon energy in the center-of-mass frame
and $s$ is a Mandelstan variable.
 At low energies, we can consider just the $S$ and $P$ waves in a fist approximation, which are described by the subscripts $l$ ($l=0$ and $l=1$), in the above expressions.  

The obtained
amplitudes are real, consequently the unitarity of the $S$ matrix
is violated. So, we unitarize the amplitudes with     
\begin{equation}
F_{l\pm}^U=\frac{F_\pm}{1-ikF_\pm}.
\end{equation}

In the center-of-mass frame 
 the differential cross sections are
\begin{equation}
\frac{d\sigma}{d\Omega}=|G|^2+|H|^2,
\label{eq:}
\end{equation}
and integrating this expression over the solid angle we obtain the total cross sections 
\begin{equation}
\sigma_T=4\pi \sum_l{\Big[(l+1)|F_{l+}^U|^2+l|F_{l-}^U|^2\Big]}
\end{equation}

\noindent
of the reactions of interest.


\section{${\bf K\Lambda}$ Interaction}
Since the $\Lambda$ has isospin 0, in
the $K\Lambda$ interaction we must consider
 $P_I=1$. For the diagram $(a)$  in Fig.1, we have, for spin-1/2 the amplitudes

\begin{figure}
\includegraphics[width=1\linewidth]{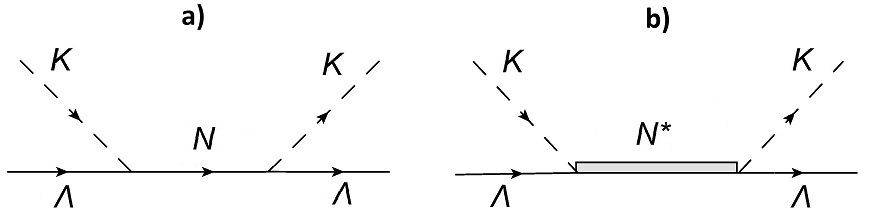}
\caption{ $K\Lambda$ Diagrams.}
\end{figure}

\begin{equation}
 A_N=\frac{g_{\Lambda KN}^2}{4m_{\Lambda}^2}(m_N+m_{\Lambda})\Big(\frac{s-m_{\Lambda}^2}{s-m_N^2}\Big),
\end{equation}
\begin{equation}
 B_N=-\frac{g_{\Lambda KN}^2}{4m_{\Lambda}^2}\bigg[\frac{2m_{\Lambda}(m_{\Lambda}+m_N)+s-m_{\Lambda}^2}{s-m_N^2}\bigg],
\end{equation}
where $m_N$ is the nucleon (or a spin-1/2 resonance) mass
 and $g_{\Lambda KN}$ are the coupling constants.

For Fig.1 $(b)$ the spin-3/2 amplitudes \cite{Olsson1975} are
\begin{equation}
A_{N^*}=\frac{g_{N^*}^2}{6}\Bigg\{\bigg[\hat{A}+\frac{3}{2}(m_\Sigma+m_{N^*})t\bigg]\bigg[\frac{2}{m_{N^*}^2-s}\bigg]
+a_0\Bigg\},
\end{equation}
\begin{equation}
B_{N^*}=\frac{g_{N^*}^2}{6}\Bigg\{\bigg[\hat{B}+\frac{3}{2}t\bigg]\bigg[\frac{2}{m_{N^*}^2-s}\bigg]-b_0\Bigg\},
\end{equation}
where
\[
\hat{A}=\frac{(m_{N^*}+m_\Lambda)^2-m_K^2}{2m_{N^*}^2}\Big[2m_{N^*}^3-2m_\Lambda^3-2m_\Lambda m_{N^*}^2-2m_\Lambda^3
\]
\begin{equation}
\label{18}
-2m_\Lambda m_{N^*}^2-2m_\Lambda^2m_{N^*}+m_K^2(2m_\Lambda-m_{N^*})\Big],
\end{equation}
\begin{eqnarray}
\hat{B}&=&\frac{1}{2m_{N^*}^2}\Big[(m_{N^*}^2-m_\Lambda^2)^2\nonumber\\
&&-2m_\Lambda m_{N^*}(m_\Lambda+m_{N^*})^2-2m_K^2(m_\Lambda+m_{N^*})^2\nonumber\\
\label{19}
&&+6m_K^2m_{N^*}(m_\Lambda+m_{N^*})+m_K^4\Big],
\end{eqnarray}
\begin{eqnarray}
\label{20}
a_0&=&-\frac{(m_\Sigma+m_\Delta)}{m_\Delta^2}(2m_\Delta^2+m_\Sigma m_\Delta-m_\Sigma ^2+2m_K^2)\nonumber\\
&&+\frac{4}{m_\Delta^2}\Big[(m_\Delta+m_\Sigma )Z+(2m_\Delta+m_\Sigma )Z^2\Big]\nonumber\\
&&\times\Big[s-m_\Sigma^2\Big],
\end{eqnarray}
\begin{eqnarray}
\label{21}
b_0&=&\frac{8}{m_\Delta^2}\Big[(m_\Sigma ^2+m_\Sigma m_\Delta-m_K^2)Z+(2m_\Sigma m_\Delta+m_\Sigma ^2)Z²\Big]\nonumber\\
&&+\frac{(m_\Sigma +m_\Delta)^2}{m_\Delta^2}+\frac{4Z^2}{m_\Delta^2}\Big[s-m_\Sigma^2\Big],
\end{eqnarray}

\noindent 
where  $m_N*$  and $m_K$ are the  spin-3/2 resonance and the kaon masses, respectively, $g_{N^*}$ are the coupling constants and $t$ is a Mandelstam variable. 

\section{${\bf K\Sigma}$ Interaction}
A similar approach  has been made in order to study the
 $K\Sigma$ interaction. The diagrams for the $N$, $N^*$ and $\Delta$ resonances
exchanges are shown in  Fig.2.
For these interactions we have different isospin projectors, $P_{\frac{1}{2}}=\frac{1}{3}\delta^{ab}+\frac{i}{3}\epsilon_{bac}\tau^c$ and $P_{\frac{3}{2}}=\frac{2}{3}\delta^{ab}-\frac{i}{3}\epsilon_{bac}\tau^c$ for the 1/2 and 3/2 isospin channels, where $a$ and $b$ are the isospin states of the $\Sigma$ hyperon.
\begin{figure}
\includegraphics[width=1.1\linewidth]{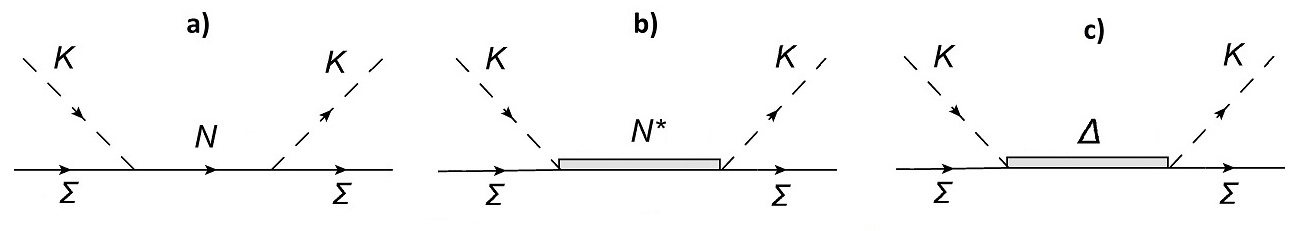}
\caption{FIG.2: $K\Sigma$ Diagrams.}
\end{figure}

For instance in the amplitudes for the $\Delta$ resonance exchange, 
considering the relation $M^\dag_b M_a=\frac{2}{3}\delta^{ab}+\frac{i}{3}\epsilon_{bac}\tau^c$,
we have 
\begin{equation}
\label{22}
A_\Delta^{+}=\frac{g_\Delta^2}{9}\Bigg\{\bigg[\hat{A}+\frac{3}{2}(m_\Sigma+m_\Delta)t\bigg]\bigg[\frac{2}{m_\Delta^2-s}\bigg]
+a_0\Bigg\},
\end{equation}
\begin{equation}
B_\Delta^{+}=\frac{g_\Delta^2}{9}\Bigg\{\bigg[\hat{B}+\frac{3}{2}t\bigg]\bigg[\frac{2}{m_\Delta^2-s}\bigg]-b_0\Bigg\},
\end{equation}
\begin{equation}
A_\Delta^{-}=\frac{g_\Delta^2}{18}\Bigg\{\bigg[\hat{A}+\frac{3}{2}(m_\Sigma+m_\Delta)t\bigg]\bigg[\frac{2}{m_\Delta^2-s}\bigg]
+a_0\Bigg\},
\end{equation}
\begin{equation}
\label{25}
B_\Delta^{-}=\frac{g_\Delta^2}{18}\Bigg\{\bigg[\hat{B}+\frac{3}{2}t\bigg]\bigg[\frac{2}{m_\Delta^2-s}\bigg]-b_0\Bigg\},
\end{equation}
where the superscript $(+)$ represents the amplitudes relative to $\delta_{ab}$ (without isospin change) and the
superscript $(-)$, the amplitudes relative to $i\epsilon_{bac}\tau^c$, that mix the isospin states. $\hat{A}$, $\hat{B}$, $a_0$ and $b_0$ in Eqs.(\ref{22}-\ref{25}) are analogous to (\ref{18}-\ref{21}) but with the replacements $m_{\Lambda}\rightarrow m_\Sigma$ and $m_{N^*}\rightarrow m_\Delta$, and $g_\Delta$ are  the coupling constants. 

The same procedure may be followed
 for the other resonances to be considered with different spin and isospin states.
There are many possible reactions that may be studied by the linear combination of the
 isospin amplitudes $T_I$ (\ref{4}),
\begin{equation}
\label{eq:}
\left\langle K^+\Sigma^+|T|K^+\Sigma^+ \right\rangle=\left\langle K^0\Sigma^-|T|K^0\Sigma^- \right\rangle=T_{\frac{3}{2}},
\end{equation}
\begin{equation}
\left\langle K^0\Sigma^+|T| K^0\Sigma^+ \right\rangle=\left\langle K^+\Sigma^-|T| K^+\Sigma^- \right\rangle=\frac{1}{3}T_{\frac{3}{2}}+\frac{2}{3}T_{\frac{1}{2}},
\label{eq:}
\end{equation}
\begin{equation}
\left\langle K^0\Sigma^0|T| K^0\Sigma^0\right\rangle=\left\langle K^+\Sigma^0|T| K^+\Sigma^0 \right\rangle=\frac{2}{3}T_{\frac{3}{2}}+\frac{1}{3}T_{\frac{1}{2}},
\label{eq:}
\end{equation}
\begin{eqnarray}
\left\langle K^0\Sigma^0|T| K^+\Sigma^- \right\rangle&=&\left\langle K^0\Sigma^+|T| K^+\Sigma^0\right\rangle
\nonumber\\
=\left\langle K^+\Sigma^-|T| K^0\Sigma^0 \right\rangle&=&\left\langle K^+\Sigma^0|T| K^0\Sigma^+\right\rangle\nonumber\\
&&=\frac{\sqrt{2}}{3}\big(T_{\frac{3}{2}}-T_{\frac{1}{2}}\big).
\end{eqnarray}
The amplitudes for each isospin channel are calculated by using the expressions  for isospin-1/2
\begin{eqnarray}
\label{eq:}
A^{\frac{1}{2}}&=&A^++2A^-,\\
\label{eq:}
B^{\frac{1}{2}}&=&B^++2B^-,
\end{eqnarray}
and for isospin-3/2
\begin{eqnarray}
\label{eq:}
A^{\frac{3}{2}}&=&A^+-A^-,\\
\label{eq:}
B^{\frac{3}{2}}&=&B^+-B^-,
\end{eqnarray}
where the last expressions are used just for the $\Delta$ resonance exchange.

\section{Conclusions}
The model for the $KY$ interactions studied in this work, 
based in non-linear Lagrangians and 
written in the partial wave formalism, presents the same simplicity
that is found in the formulation of the $\pi N$ and $\pi Y$ interactions 
\cite{BarrosJr2001}. 

In a further paper we will show the complete results of the observables, cross sections,  
polarizations and phase-shifts of the low energy $KY$ interactions considering also the exchange of $\rho$ and $\sigma$ mesons and the crossed diagrams \cite{Marcelo2018}, not
presented in this work. 

Another result of interest is the $D$-wave phase-shift
for the $\overline K\Lambda$ interaction at the $\Omega$ baryon mass,  that may be used in 
the study of the $CP$ violation \cite{BarrosJr2003} in the $\overline K\Lambda\rightarrow\Omega$ decay.

\end{document}